\documentclass[prd,superscriptaddress,nofootinbib]{revtex4}

\usepackage{amsmath}
\usepackage{graphicx}
\usepackage{epstopdf}
\usepackage{verbatim}
\newcommand{\be}{\begin{equation}}
\newcommand{\ee}{\end{equation}}
\newcommand{\bea}{\begin{eqnarray}}
\newcommand{\eea}{\end{eqnarray}}

\begin{document}
\title{Gravitational collapse of K-essence Matter in Painlev\'{e}-Gullstrand coordinates}

\author{C. Danielle Leonard}
\affiliation{Department of Physics and Astronomy, University of Waterloo, Waterloo, Ontario, N2L 3G1, Canada}
\affiliation{Department of Physics and Physical Oceanography, Memorial University of Newfoundland, St. John's, Newfoundland, A1B 3X7, Canada}
 
\author{Jonathan Ziprick}
\affiliation{Perimeter Institute, 31 Caroline Street North, Waterloo, Ontario, N2L 2Y5, Canada}
\affiliation{Department of Physics and Astronomy, University of Waterloo, Waterloo, Ontario, N2L 3G1, Canada}

\author{Gabor Kunstatter}
\affiliation{Department of Physics and Winnipeg Institute of Theoretical Physics, University of Winnipeg, Winnipeg, Manitoba, R3B 2E9, Canada}

\author{Robert B. Mann}
\affiliation{Department of Physics and Astronomy, University of Waterloo, Waterloo, Ontario, N2L 3G1, Canada}
\affiliation{Perimeter Institute, 31 Caroline Street North, Waterloo, Ontario, N2L 2Y5, Canada}

\date{\today}

\begin{abstract}
We conduct numerical simulations in Painlev\'{e}-Gullstrand coordinates of a variety of K-essence-type scalar fields under spherically symmetric gravitational collapse. We write down generic conditions on the K-essence lagrangian that can be used to determine whether superluminality and Cauchy breakdown are possible. Consistent with these conditions, for specific choices of K-essence-type fields we verify the presence of superluminality during collapse, while for other type we do not.  We also demonstrate that certain choices of K-essence scalar fields present issues under gravitational collapse in Painlev\'{e}-Gullstrand coordinates, such as a breakdown of the Cauchy problem.  
\end{abstract}


\maketitle

\section{Introduction}

In recent years, there have been a number of models put forward to explain the accelerating expansion of the universe.  Falling under the general rubric of dark energy, such models 
have included quintessence \cite{quin}, Elko spinors \cite{Elko}, H-essence  \cite{hess}, quintom matter \cite{quintom}, phantom matter \cite{phantom} and more. 
 However, these models do not explain why the expansion of the universe is accelerating while humans are here to observe it (the cosmic coincidence problem). Models that avoid anthropic rationales for explaining cosmic coincidence generally require the fine-tuning of parameters.  It was with the intention of avoiding both fine-tuning and anthropic arguments that K-essence \cite{dynamic,essentials} was first proposed.  

K-essence models invoke dark energy in the form of a scalar field with non-linear kinetic energy.  These models are constructed so that the scalar field develops a negative pressure once the matter dominated era begins, and so
the associated dynamical behaviour is then deemed responsible for the accelerating cosmic expansion of
our universe.  
K-essence initially came under criticism \cite{nogo} because it was shown that any such model that solved the coincidence problem also necessarily involved perturbations propagating faster than light.  However, the proposers of K-essence defended their model \cite{superlum} by demonstrating that despite superluminality, causality was preserved for physical situations.
 
It is due to this intrinsic superluminality that the idea of gravitational collapse of K-essence matter has become recently popularized.  A number of K-essence-type models, characterized by non-linear kinetic energy, have emerged.  These models serve the primary purpose not of solving the coincidence problem, but of exploring the idea of superluminal information escaping from a black hole.  It was shown initially in \cite{escape} that for the stationary black hole case, signals could escape from within the light horizon as superluminal perturbations.  More recently, a dynamic collapse scenario was considered and numerically simulated in maximal slicing coordinates \cite{MS}.  This study found evidence of superluminality in certain Lagrangian choices for K-essence. Other choices of Lagrangian, however, showed no such evidence.  This study also demonstrated that in some choices of Lagrangian we are limited by the eventual breakdown of the Cauchy problem.  
 
Because the well-posedness of the Cauchy problem is dependent on the time-coordinate remaining globally valid, the authors of \cite{MS} were unable to determine whether the observed breakdown was a result of the choice of Lagrangian or of coordinate systems.  To this end, we now examine a selection of choices of Lagrangians in Painlev\'{e}-Gullstrand (P-G) coordinates, which are regular across future luminal horizons. We provide general analytic expressions for determining whether superliminal propagation and/or Cauchy breakdown occur for specific classes of K-essence fields. Because P-G coordinates offer an indisputably distinct time coordinate from maximal slicing coordinates, simulations in P-G should illuminate the coordinate dependence of Cauchy breakdown or lack thereof. Using P-G coordinates is also a strong check to ensure the coordinate independence of the results of the maximal slicing simulations in general.

Through a series of numerical simulations in P-G coordinates of three models of K-essence, we demonstrate that the gravitational collapse of K-essence can produce either superluminal or subluminal sonic horizons.  Hence we verify the qualitative results of \cite{MS} and therefore provide evidence for the coordinate independence of these results.  We further examine the breakdown (or lack thereof) of the Cauchy problem for these three models.  We find that one model experiences a breakdown of this problem in many parameter choices, and that the other two models do not experience Cauchy breakdown.  This corroborates the results found in \cite{MS}.  Therefore, our results are indicative that the issue of Cauchy breakdown for these models of K-essence is coordinate independent.

\section{Methods}

We start with a general action for K-essence coupled to gravity, given by
\begin{equation}\label{action}
I = \int {\sqrt g} \, {d^4} x \, \left ( {\frac {{{}^{(4)}}R} {16\pi G}} + {\cal L}(X) \right )
\end{equation}
where ${{}^{(4)}}R$ is the spacetime scalar curvature, $G$ is Newton's gravitational constant in four dimensions, and 
$ X = - {\frac 1 2} {\nabla ^a} \psi {\nabla _a} \psi$ where $\psi$ is the K-essence scalar field.  

\medskip

If we vary the action with respect to the metric, we obtain the Einstein field equation
\begin{equation}
{G_{ab}} = \kappa {T_{ab}}
\label{efe}
\end{equation}
where we have denoted denoted the Einstein tensor as $G_{ab}$, and we note that for K-essence we have
\begin{equation}
{T_{ab}} =  {{\cal L}_X}  {\nabla _a}\psi {\nabla _b} \psi + {\cal L} {g_{ab}}
\label{stress tensor}
\end{equation}
Above and throughout this paper we use ${\cal L}_X$ to mean $d{\cal L}/dX$ and
${\cal L}_{XX}$ to mean ${d^2}{\cal L}/d{X^2}$.
We vary the action with respect to $\psi$, the K-essence scalar field, and obtain equations of motion for K-essence:
\begin{equation}
{{\tilde g}^{ab}} {\nabla _a}{\nabla _b} \psi = 0
\label{kefe}
\end{equation}
Note that in the above we have used ${{\tilde g}^{ab}}$, which is the effective (inverse) metric governing propagations of perturbations of the K-essence field, given by:
\begin{equation}
{{\tilde g}^{ab}} = {{\cal L}_X}  {g^{ab}} - {{\cal L}_{XX}} {\nabla ^a}\psi {\nabla ^b}\psi
\label{geff}
\Longrightarrow
{{\tilde g}_{ab}} = \frac{1}{{\cal L}_X}  {g_{ab}} + c_s^2 \frac{{\cal L}_{XX}}{{\cal L}^2_X} {\nabla_a}\psi {\nabla_b}\psi
\end{equation}
where
$$
c^2_s = \frac{{\cal L}_X }{{\cal L}_X +2X{\cal L}_{XX}} 
$$
is the speed of sound.

The luminal apparent horizon forms when surfaces of constant $r$ become null, implying
\begin{equation}
\label{lumhor1}
g^{ab}\nabla_a r \nabla_b r=0 \Longrightarrow g^{rr}=0 
\end{equation}
The right hand side assumes only a coordinate basis with $r$ as the spatial coordinate. 

We are now able to examine general conditions for superluminal propagation and Cauchy breakdown.
The condition for the formation of the sonic horizon requires that the same general condition as (\ref{lumhor1}) be satisfied for the effective metric of K-essence, that is 
\begin{equation}
\label{sonhor1a}
{\tilde g}^{ab}\nabla_a r \nabla_b r= 0
\end{equation}
which, in analogy to the luminal case leads to:
\begin{equation}
{\tilde g}^{rr}=0 
\end{equation}
A  calculation reveals that in general
\be
{{\tilde g}^{rr}} = {{\cal L}_X}  {g^{rr}} - {{\cal L}_{XX}} \left({\nabla ^r}\psi\right)^2 
   = g^{rr}\left(\mathcal{L}_{X}+2X\mathcal{L}_{XX}\right)- \mathcal{L}_{XX}
   \frac{r^2}{(-4gl^{2})}  (\dot{\psi})^2=
   \frac{{\cal L}_X}{c^2_s}   g^{rr}  - \mathcal{L}_{XX}  \frac{r^2}{(-4gl^{2})}   (\dot{\psi})^2
    = 0
   \label{tilde grr}
\ee
where in the last line $g$ is the determinant of the 2-metric as given in equation (\ref{met2d}), and $l$ is an arbitrary length scale.  Note that $g$ is negative, so $-g$ is positive.  Eq.(\ref{tilde grr}) shows that the relative sign difference between the sonic horizon condition and the luminal horizon condition is determined by the signs of $\mathcal{L}_X$ and $\mathcal{L}_{XX}$. Superluminal escape   requires the sign of $\tilde{g}^{rr}$ to be positive  just inside a black hole horizon, where $g^{rr}$ is negative. We assume that $\mathcal{L}_X$ is positive, which is necessary for the kinetic term in the stress energy tensor to be positive (cf. Eq.(\ref{stress tensor})). Thus, for the right hand side of equation (\ref{tilde grr}) to be positive, it is necessary (but not sufficient) that $\mathcal{L}_{XX}<0$.  When this condition is satisfied, we can generically say that superluminal escape is possible.  This will be verified in the numerical studies below.    

Another key aspect of our method is the examination of the potential breakdown of the Cauchy problem. 
A breakdown of the Cauchy problem occurs when the surfaces of constant time 
 (\ref{geff}) become null with respect to the effective metric.  We can write this as  
 \begin{equation}
\label{cauchy1}
{\tilde g}^{ab}\nabla_a t \nabla_b t= 0 \Longrightarrow  {\tilde g}^{tt}=0 
\end{equation}
which using (\ref{psidot}), may be suggestively rewritten as 
\begin{eqnarray}
{\tilde g}^{tt}&=&{{\cal L}_X}  {g^{tt}} - {{\cal L}_{XX}} \left({\nabla ^t}\psi\right)^2  
= \frac{{\cal L}_X}{c^2_s}   g^{tt}  - \mathcal{L}_{XX}  \frac{r^2}{(-4gl^{2})}   ({\psi}^\prime)^2
   \label{tilde gtt}
\end{eqnarray}
with $g$ and $l$ defined as in equation (\ref{tilde grr}).
In simulating equation (\ref{efe}), we require that surfaces of constant time be spacelike; in other words we require that $g^{tt}<0$.  However, for K-essence  we must then require that in addition to this, the surfaces of constant time of the effective metric are also spacelike. That is, ${{\tilde g}^{tt}}<0$.  Clearly, for certain Lagrangian choices, ${{\tilde g}^{tt}}$ can go to zero independently of $g^{tt}$.  When surfaces of constant time are no longer spacelike with respect to either the background or effective metric, we say that the Cauchy problem ceases to be well posed on this surface, and our simulations must halt.  
Eq.(\ref{tilde gtt}) and Eq.(\ref{tilde grr}) show that   the condition for Cauchy breakdown in a theory is the same as those for superluminal propagation, though the onset of when each occurs depends on the relative magnitude of $|g^{rr}/g^{tt}|$.

Of course the above assumes that both $\mathcal{L}_{X}$ and $\mathcal{L}_{XX}$ are everywhere regular. It is also possible for pathologies to occur if either becomes singular. This too will be illustrated in one of the examples to follow.
 
In order to be able to distinguish physical properties of the K-essence field from a breakdown of the coordinates it is important to use spatial slicings that are regular at luminal horizons. To this end
 we work in Painlev\'{e}-Gullstrand coordinates.  Our simulations are a generalization to the K-essence case of the method developed in \cite{jon}.  
In Painlev\'{e}-Gullstrand coordinates the metric take the form:
\begin{equation}
ds^2= -\sigma^2dt^2+\left(dr + \sqrt{\frac{2G\mathcal{M}}{r}}\sigma dt\right)^2 + r^2 d\Omega^2 
\label{PG metric}
\end{equation}
where in the dynamical case both the lapse function $\sigma$ and the Misner-Sharpe mass function $\mathcal{M}$ are functions of both $r$ and $t$. Note that spatial slices ($dt=0$) are flat so that  P-G coordinates are singular only at the origin, $r=0$ as long as $\sigma^2$ and $\cal M$ are positive.  This means that we are able to continue to evolve the simulation inside any horizons that form (as opposed to the situation in Schwarzschild coordinates, for example).  Another reason for our choice of coordinates is simply to avoid issues associated with the usual choice of polar-radial coordinates.  This choice would be inappropriate in this setting, because such coordinates are only valid until the first trapped surface forms.  In this case, since we have superluminality potentially present, K-essence is not trapped by normal trapped surfaces, but rather by the trapped surfaces of the effective metric. By working in P-G coordinates we instead can follow the evolution of the system much further, providing a different qualitative perspective from that of  maximal slicing coordinates, as noted above.
 
\begin{figure}
\includegraphics[scale=0.25]{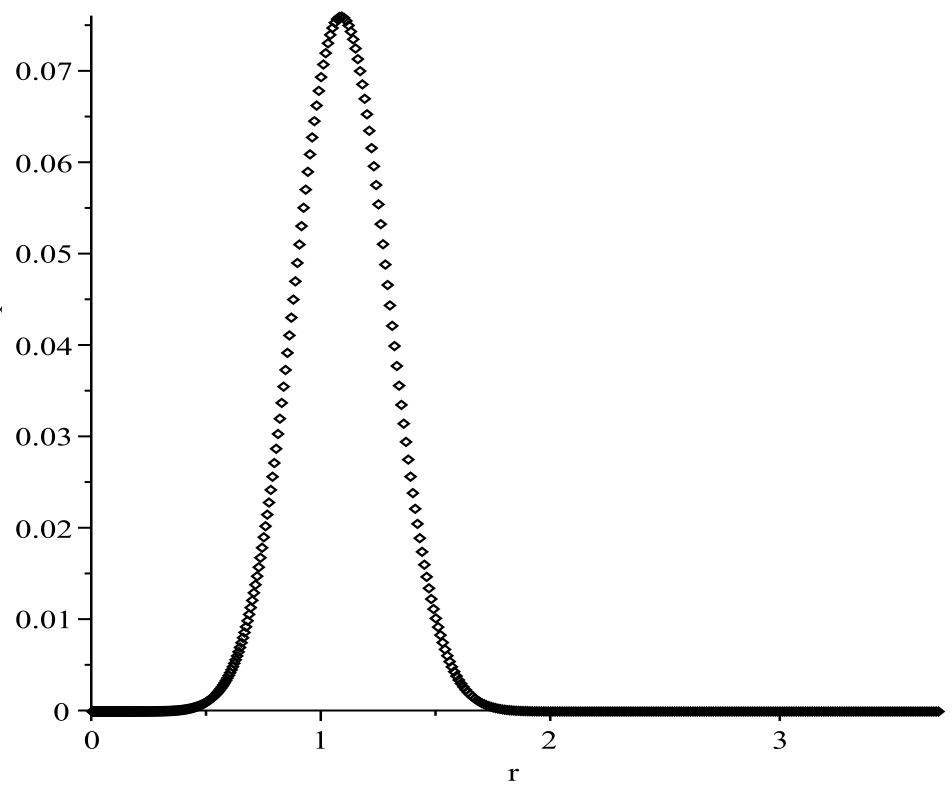}
\caption{\label{initpsifig}$\psi$ {\it vs} $r$ at $t=0$}
\end{figure}

Since we are considering spherically symmetric collapse, the number of dynamical degrees of freedom is reduced to that of $\psi$ and its conjugate momentum $\Pi_{\psi}$.  These quantities obey the equations (\ref{psidot},\ref{Pipsidot}) given in appendix A, which contains  more detailed description of our dynamical variables and specific numerical methods.  For now, it will suffice to note our initial conditions on $\psi$ and $\Pi_{\psi}$, which we take to be
\begin{equation}
\psi(t=0)=Ar^{2}\exp\Bigg(-\frac{(r-r_{0})^{2}}{B^{2}}\Bigg)
\label{initpsi}
\end{equation}
as illustrated in fig \ref{initpsifig} and 
\begin{equation}
\Pi_{\psi}(t=0)=0
\label{initPi}
\end{equation}
for $\Pi_{\psi}$.
For all of our simulations, we chose $B=0.3$ and $r_{0}=1.0$.  

\section{Results}

Before proceeding to the results for each specific choice of Lagrangian, we review the behaviour of a spherically symmetric gravitational collapse scenario for `normal matter' (where ${\cal L}(X) = X$) and note the conditions for the formation of light apparent horizons and sonic horizons (horizons trapping the perturbations of the effective metric) in P-G coordinates.

Let us recall that for initial data such as that we have chosen, there are a number of different possibilities for how the collapse of normal matter proceeds.  If the amplitude of the scalar field, A, as given by equation (\ref{initpsi}) is sufficiently small, subcritical collapse occurs.  This means that no horizons are formed, and the ingoing pulse becomes, at the origin, an outgoing pulse and the field simply `passes through' itself.  However, if A exceeds some critical value, the collapse becomes supercritical.  Luminal apparent horizons form, from inside which nothing can escape.

K-essence matter is governed by the effective metric $\tilde{g}$, and as shown above perturbations in this matter are permitted to exceed the speed of light.  In P-G coordinates, the two distinct conditions for the formation of the luminal and sonic horizons, are respectively  
\begin{equation}
\label{ahcond}
r-2G\mathcal{M}=0
\end{equation}
and
\begin{equation}
\left(1-\frac{2G\mathcal{M}}{r}\right)\left({\mathcal L}_X + 2X {\mathcal L}_{XX}\right)-{\mathcal L}_{XX}\left(\frac{\Pi_\psi}{4\pi r^2 {\mathcal L}_X} + \sqrt{\frac{2G\mathcal{M}}{r}} \psi^\prime \right)^2 = 0
\label{shcond2}
\end{equation}

Similarly we can find a general condition for the breakdown of the Cauchy problem in P-G coordinates.  By noting that this occurs when surfaces of constant $t$ are no longer timelike (as stated above), we obtain the general condition 
\begin{equation}
\label{cauchy}
{\cal L}_X +  \left(\frac{\Pi_\psi}{4\pi r^2 {\cal L}_X}\right)^2{\cal L}_{XX} = 0
\end{equation}

Eqs.(\ref{ahcond},\ref{shcond2},\ref{cauchy}) are used in the numerical code to  locate luminal horizons, sonic horizons and Cauchy breakdown, respectively.

\subsection{Born-Infeld Lagrangians}

We examine two choices of Born-Infeld Lagrangians.  These are based on the Lagrangian given in \cite{escape}, which was proposed to demonstrate information escaping from inside a stationary luminal apparent horizon.  As such they represent interesting models for studying gravitational collapse since it is not immediately clear what effect the superluminal character of collapsing K-essence matter will have on this process.

The first of this type of Lagrangian is given by 
\begin{equation}
\mathcal{L}_{1}(X)=\alpha^{2}[\sqrt{1+\frac{2X}{\alpha^{2}}}-1]-\Lambda
\label{sh1}
\end{equation}
where $\alpha$ is the K-essence parameter.  As $\alpha$ goes to $\infty$, we recover $\mathcal{L}(X)=X$, the standard Lagrangian for a massless scalar field (`normal' matter). We choose $\Lambda=0$ for our simulations.  
Note that for gravitational collapse it is possible for $X$ to become negative,
unlike cosmological scenarios where homogeneity implies all physical quantities depend solely on time ensuring $X>0$.  
\begin{figure}
\includegraphics[scale=0.4]{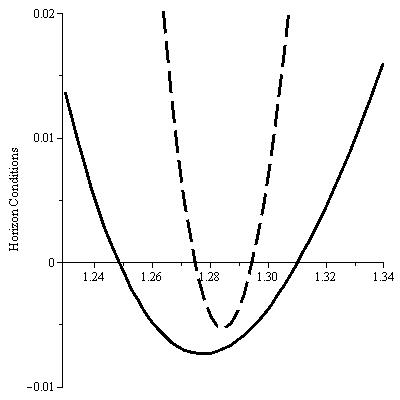}
\caption{\label{splum}Sonic (dashed) and light (solid) horizon formation conditions vs $r$ at P-G time $t=0.172499478$ for $\mathcal{L}_{1}(X)$ with A=0.13 and $\alpha=3.0$}
\end{figure}

\begin{figure}
\includegraphics[scale=0.2]{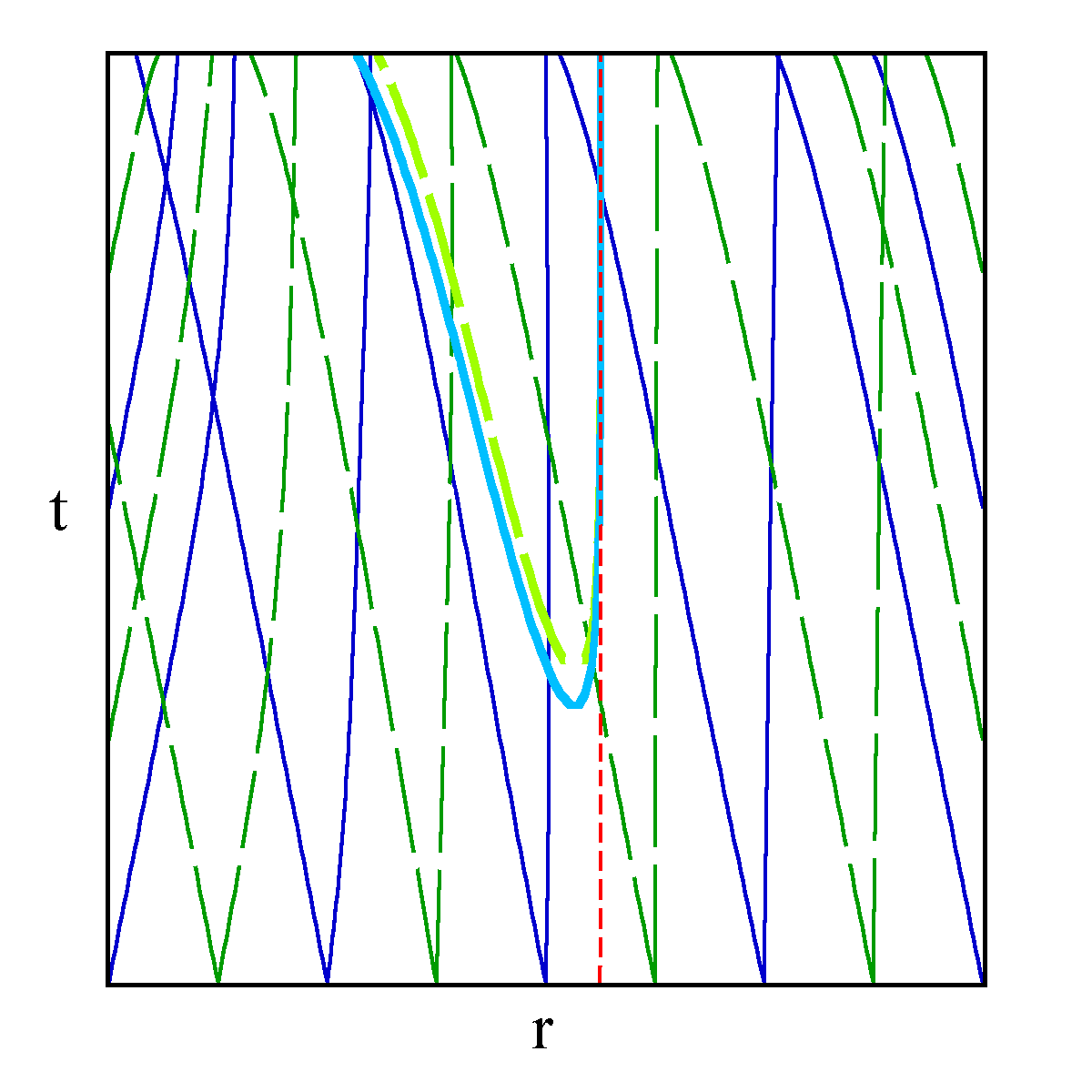}
\caption{\label{l1nullmap} Map of null geodesics for $\mathcal{L}_{1}(X)$ of the background metric (dark blue solid) and the effective k-essence metric (dark green dashed) as well as the luminal apparent horizon (light blue solid) and sonic apparent horizon (light green dashed), for parameters A=0.12, $\alpha$=2.0. The red line is the sonic event horizon, i.e. the last outgoing sound wave that does not quite escape the sonic trapping region. The luminal event horizon is just slightly to the right of the sonic event horizon, but is too close to distinguish on this scale. }
\end{figure}

{}From figure \ref{splum} we see that for the Lagrangian given in (\ref{sh1}), we have the possibility of superluminal escape from the black hole.  For the given choice of parameters, we clearly see that the sonic horizon forms inside the apparent horizon, indicating that perturbations in the intermediate region which travel faster than light can escape the light horizon.  The spatial distinction between the two horizons does not persist for long though: by P-G time $t=0.199998676$, the two horizons have merged spatially within the precision of our numerics.  This behaviour was also qualitatively observed in maximal slicing coordinates in \cite{MS}.  As described therein, the reason for the rapid merger of the sonic and apparent horizons is that the K-essence scalar field quickly falls into the black hole or propagates outwards away from the horizon.  This leaves the background and effective metrics equivalent in the horizon region, so luminal and sonic horizons are both dependent only on the background metric.   This behaviour is equivalently demonstrated in figure \ref{l1nullmap}, which depicts the null geodesics of both the background and effective metrics, as well as both the sonic and luminal apparent horizons for a typical choice of parameters. We find, somewhat surprisingly, that the inner sonic and luminal trapping horizons remain separated, while the outer ones merge very rapidly, leaving only a very narrow region of superluminal escape. 

Aside from the striking example provided in figure \ref{splum}, it will be instructive to provide a more in-depth discussion of the behaviour of this model in the parameter space explored.  This will also serve as a discussion of breakdown of the Cauchy problem in this model.  We examined this choice of $\mathcal{L}$ for a range of parameters: our values of A ran from 0.01 to 0.15, and our values of $\alpha$ from 0.05 to 400.    Within this range, there were a number of different categories of behaviour observed.  In general terms, all parameter pairing for this choice of $\mathcal{L}$ may be grouped into one of the following categories:

\begin{enumerate}
\item{Subcritical Collapse - no Cauchy breakdown}
\item{Cauchy breakdown prior to any horizons forming}
\item{Cauchy breakdown after apparent horizon formation but before sonic horizon formation}
\item{Cauchy breakdown after both horizons form, where sonic horizon appears inside light horizon}
\item{Cauchy breakdown after both horizons form, where within our numerical precision, sonic and apparent horizons are indistinguishable}
\end{enumerate}

Before discussing specifically for which parameters each behaviour is present, we note the factors which could create different behaviours in different parts of the parameter space.  First, we know that as $\alpha$ increases we approach the behaviour of `normal' matter, and as $\alpha$ decreases we depart from this behaviour.  Therefore, we expect breakdown of the Cauchy problem to occur sooner for cases of smaller $\alpha$.  Secondly, we know that as A increases, the collapse will go from from subcritical to supercritical.  Within the supercritical range, as A becomes bigger, collapse will be more rapid and horizon formation will occur sooner.  With this in mind, we summarize our observations for $\mathcal{L}_{1}$ as follows.

\begin{figure}
\includegraphics[scale=0.4]{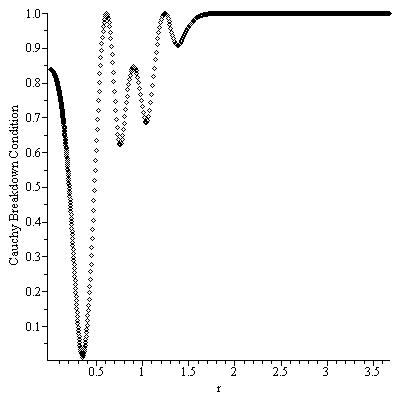} 
\caption{\label{cauchyfigure} Cauchy breakdown condition vs $r$ for $\mathcal{L}_{1}$ at P-G time t=0.497767983 (the last timestep before Cauchy breakdown) for A=0.1 and $\alpha=0.5$.  The point of closest approach to zero is approximately r=0.3434.}
\end{figure}

\begin{enumerate}
\item{Subcritical collapse without Cauchy breakdown occured for $A=0.01-0.04$ for relatively large $\alpha$ - the minimal value being between 0.01 and 2.0, depending on the value of $A$, and extending to $\alpha=400.0$ in all cases.  This is as we expect - large $\alpha$ implies normal matter behaviour, and small A in normal matter corresponds to subcritical collapse.}
\item{The breakdown of the Cauchy problem prior to any horizon formation occured at small $\alpha$ for all values of $A$.  Figure \ref{cauchyfigure} shows an example of this type of behaviour. Depending on the value of $A$, the maximum value of $\alpha$ for which this occurred was between 0.075 and 80.0.  This is explained by the fact that, as stated before, at small $\alpha$ we have behaviour less typical of normal matter and we expect Cauchy breakdown to occur sooner.}
\item{We observed 2 cases of  Cauchy breakdown occurring between the formation of the sonic horizon and the formation of the luminal apparent horizon.  This was for $A=0.11$ and $\alpha=0.5,0.75$.  This behaviour is understood easily once we understand that in this choice of $\mathcal{L}$ sonic perturbations can be superluminal - the  luminal apparent horizon would then of course form before the sonic apparent horizon.  In these cases, Cauchy breakdown occurs between these formations.}
\item{For intermediate values of $\alpha$ and large values of $A$, we observe  breakdown of the Cauchy problem after luminal and sonic horizons have formed, with the sonic horizon initially occurring inside the luminal horizon.   This type of collapse was observed for $A=0.1-0.13$ and for values of $\alpha$ ranging from 0.5 to 25.0 depending on the value of $A$.  As we increase $\alpha$ within this range, the window for superluminal escape becomes smaller in time, or in other words the two horizons converge  more rapidly. This is as expected, since as $\alpha$ increases we progress towards normal matter behaviour, so Cauchy breakdown occurs later and  the two horizons become indistinguishable. The two horizons also converge more rapidly as $A$ is increased, because increasing $A$ increases the rapidity of the collapse.}
\item{For large values of $\alpha$ and large values of $A$, we find that Cauchy breakdown occurs after both horizons have formed. However we are unable to numerically distinguish the locations of the horizons to within
our levels of tolerance.  Again, this is as expected 
   because for   normal scalar field collapse there is no distinction between the luminal and sonic horizons and
   large $\alpha$ is
     the normal matter limit. }
\end{enumerate}

Summarizing, for this choice of $\mathcal{L}$ we observe that superluminal escape is possible for a portion of the parameter range.  To observe this, we must have $\alpha$ sufficiently large to allow us to observe horizons prior to Cauchy breakdown, and we must also have $\alpha$ sufficiently small so that that the two horizons are numerically distinguishable.  We also require $A$ large enough to avoid subcritical collapse, but small enough to avoid the two horizons converging too rapidly to be distinguished numerically.

We also see for this choice of $\mathcal{L}$ that we have Cauchy breakdown occuring at small values of $\alpha$ for all $A$, in   accord with results for the MS case \cite{MS}.  That we observe this breakdown for similar regions of the parameter space using an entirely different time coordinate indicates that the breakdown of the Cauchy problem in this choice of Lagrangian is likely a generic coordinate-independent feature.  Conversely, for $A=0.05-0.13$, the breakdown of the Cauchy problem occurred eventually for all values of $\alpha$ which we examined.   Therefore we see that in P-G coordinates the breakdown of the Cauchy problem is not restricted to small $\alpha$ as it is in maximal slicing coordinates.  

\medskip

We can easily modify the Lagrangian (\ref{sh1}) to avoid the problem of Cauchy breakdown by choosing a second Born-Infeld type Lagrangian \cite{MS} \begin{equation}
\mathcal{L}_{2}(X)=\alpha^{2}[1-\sqrt{1-\frac{2X}{\alpha^{2}}}]-\Lambda
\label{sh2}
\end{equation}
where again taking the limit as $\alpha$ goes to $\infty$ yields normal matter.  
\begin{figure}
\includegraphics[scale=0.4]{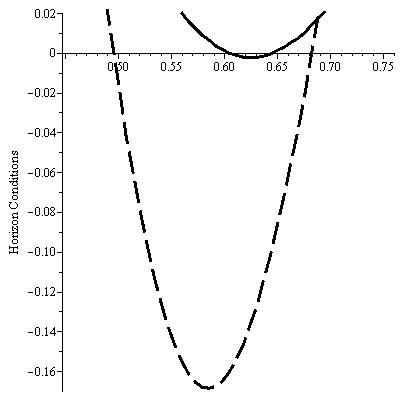}
\caption{\label{sublum} Sonic (dashed) and light (solid) horizon formation conditions vs $r$ at P-G time   $t=0.744807261$ for $\mathcal{L}_{2}(X)$ with A=0.1 and $\alpha=2.0$}
\end{figure}

\begin{figure}
\includegraphics[scale=0.2]{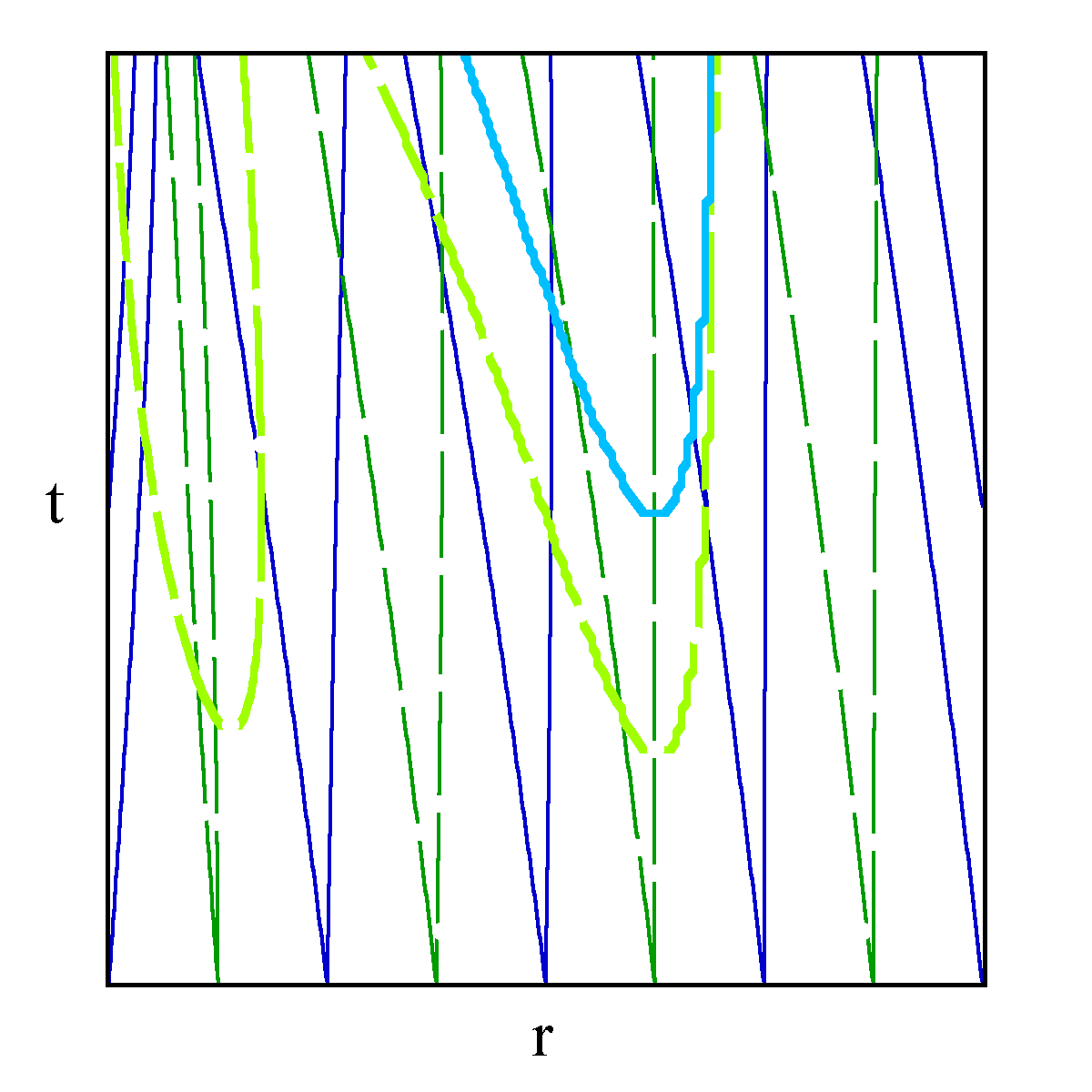}
\caption{\label{l2nullmap} Map of null geodesics $\mathcal{L}_{2}(X)$ of the background metric (dark blue solid) and the effective k-essence metric (dark green dashed) as well as the luminal apparent horizon (light blue solid) and sonic apparent horizon (light green dashed), for parameters A=0.1, $\alpha$=2.0}
\label{nullmap L2}
\end{figure}

Setting  $\Lambda$  to 0, in this case  we find no evidence in our simulations of superluminal behaviour.  Our range of parameters observed in this case was the same as for $\mathcal{L}_{1}$.  For a typical choice of parameters we see in figure  \ref{sublum}  that the sonic horizon forms outside of the light horizon, as we would expect for normal matter.  Hence, there is no superluminal propagation in this case.   This is further illustrated in figure \ref{l2nullmap} by mapping null geodesics and horizons for both background and effective metrics.   As for $\mathcal{L}_{1}$, we observed the two outer trapping horizons merging quickly, for the same reason, and the inner ones remaining separated.   For the parameter values used in figure  \ref{sublum}, the two horizons were indistinguishable by P-G time t=0.779575347.  

While we did not see any Cauchy breakdown in this case, we did encounter another issue with this choice.  As can be clearly seen by the form of equation (\ref{sh2}), it is possible for $\mathcal{L}_{X}=\frac{1}{\sqrt{1-\frac{2X}{\alpha^{2}}}}$ to become infinite when $X=\frac{\alpha^{2}}{2}$.  However, \cite{superlum} clearly states that $\mathcal{L}_{X}$ should not be allowed to become infinite. Our simulations indeed become numerically unstable in cases where X evolves to this value.

In the MS case \cite{MS}, simulations were reported to eventually fail not for this reason, but due to the formation of shock waves of perturbations.  This is indicated by the values of the speed of sound  $c_{s} $  going to 0, where 
\begin{equation}
c_{s}^{2}= \frac{\mathcal{L}_{X}}{\mathcal{L}_{X}+2X\mathcal{L}_{XX}} = 1-\frac{2X}{\alpha^{2}}
\label{cs}
\end{equation}
and so $c_s\to 0$   as X approaches $\frac{\alpha^{2}}{2}$.   

The full description of behaviour for each portion of the parameter space we examined is analogous to that for $\mathcal{L}_{1}$ in many ways, so we present a slightly abbreviated version of our results compared to $\mathcal{L}_{1}$.  
\begin{enumerate} 
\item{Subcritical collapse: as before, this occurred for small A and relatively large $\alpha$, consistent with the limit of normal matter.}
\item{We found that $\mathcal{L}_{X}$ could become infinite after the formation of only a sonic horizon.  This is somewhat analogous to case 3 for $\mathcal{L}_{1}$, in which Cauchy breakdown occurred after the formation of only a luminal horizon.   In this case, $\mathcal{L}_{X}$  becomes infinite at small values of $\alpha$ for all values of $A$.  We note that in reality this case encompasses two distinct cases.  In one case, we have this occuring for values of A which at larger $\alpha$ produce both horizons prior to $\mathcal{L}_{X}$ becoming infinite (see below).  This case is not unexpected: we simply see $\mathcal{L}_{X}$ diverging prior to the expected luminal horizon forming.  The second case is more thought provoking: this behaviour also occurs for values of A for which at larger $\alpha$ we see subcritical collapse with no horizons of any type forming (see above). This indicates there must be a threshold value of $\alpha$ below which subcritical behaviour yields to the formation of a sonic horizon.  This behaviour is unexpected and may merit further investigation.}
\item{For mid-range values of $\alpha$ and large values of $A$ we found that $\mathcal{L}_{X}$ became infinite after both horizons formed, with the sonic horizon appearing outside the light horizon. As before the two horizons were visibly separated initially, then converged together. }
\item{Finally, for large $\alpha$ and large $A$, we again found both horizons forming prior to $\mathcal{L}_{X}$ becoming infinite. We were  unable to distinguish between the two horizons within our numerical limits of tolerance. }
\end{enumerate}

As before, we observed  larger $A$ to imply more rapid gravitational collapse and hence  horizon formation taking place sooner. While we found, as expected, no Cauchy breakdown in the parameter space for this model, we
encountered a new problem: that of 
  diverging $\mathcal{L}_{X}$.   Specifically, we found that $\mathcal{L}_{X}$  became infinite more rapidly for smaller values of $\alpha$, consistent with increasing departure from the normal matter limit.  Our results are commensurate with the MS case \cite{MS} and indicate  that the Cauchy problem (or lack thereof) and  the problem of diverging $\mathcal{L}_{X}$ (or vanishing $c_s$)  are likely coordinate independent.

\subsection{Lagrangians Allowing a Global Time Coordinate}

We next examine the behaviour of a choice of Lagrangian that is completely free from the problem of Cauchy breakdown by construction.  Such a Lagrangian would have the left hand side of  (\ref{cauchy}) be either positive everywhere or negative eveywhere.  One such example  is \cite{MS}
\begin{equation}
\mathcal{L}_{3}(X)=\frac{CX}{1+C}+e^{\frac{X}{1+C}}-1
\label{Lexp}
\end{equation}
where C is a parameter of the model; as C goes to $\infty$ we recover the behaviour of normal matter ($\mathcal{L}=X$).  We can verify quite easily using a more general form of (\ref{cauchy}) that this makes the left hand side of (\ref{cauchy}) positive everywhere, therefore entirely avoiding Cauchy breakdown.  
\begin{figure}
\includegraphics[scale=0.4]{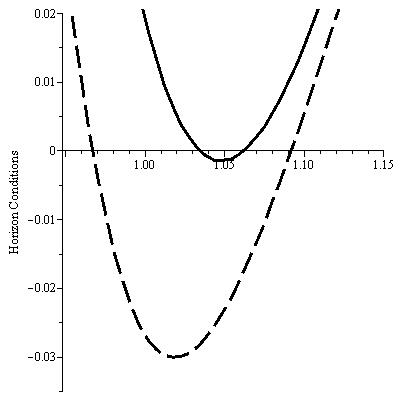}
\caption{\label{lexpf}  Sonic (dashed) and light (solid) horizon formation conditions vs r at P-G time   $t=0.399778586$ for $\mathcal{L}_{3}(X)$ with A=0.12 and C=0.8}
\end{figure}

\begin{figure}
\includegraphics[scale=0.2]{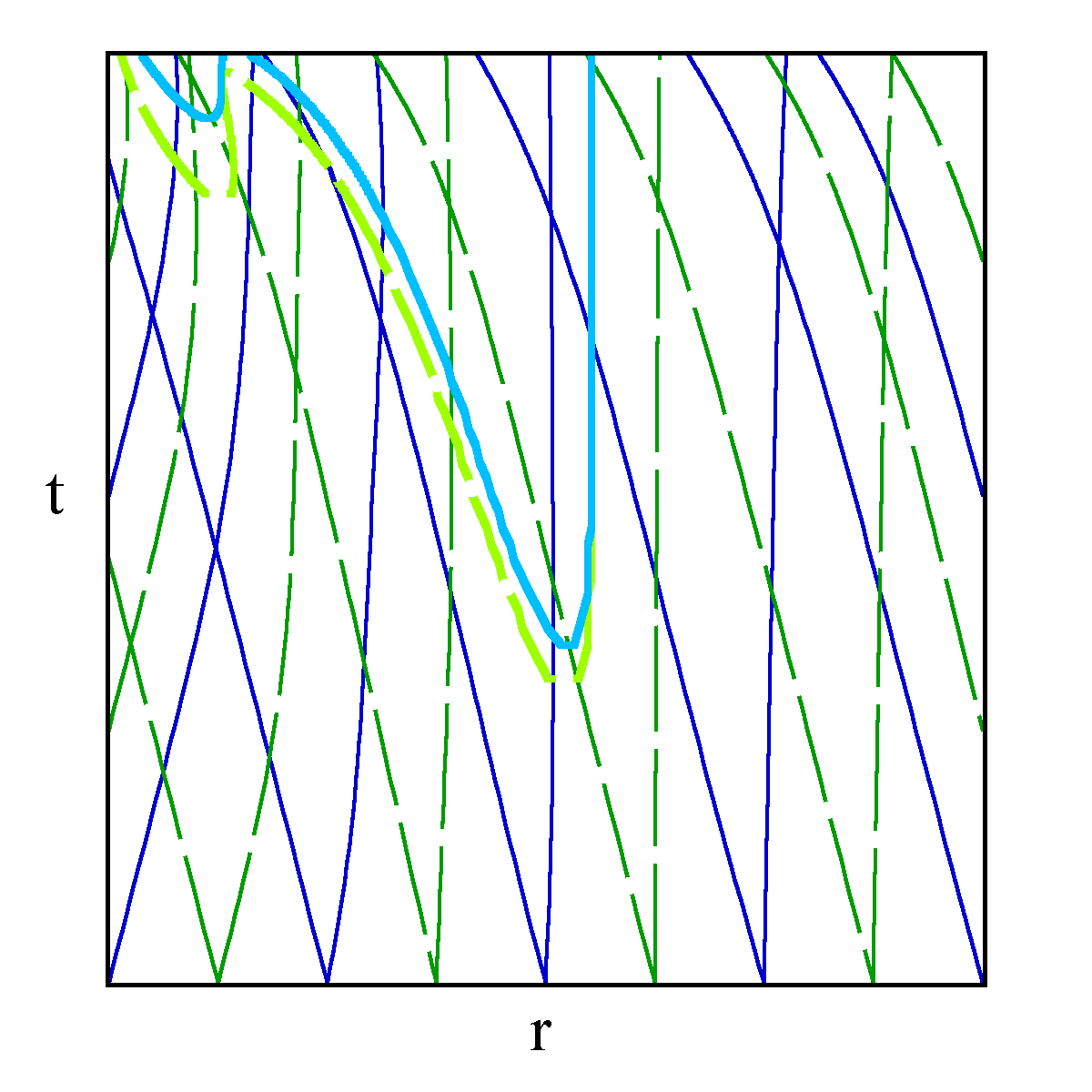}
\caption{\label{l3nullmap} Map of null geodesics $\mathcal{L}_{3}(X)$ of the background metric (dark blue solid) and the effective k-essence metric (dark green dashed) as well as the luminal apparent horizon (light blue solid) and sonic apparent horizon (light green dashed), for parameters A=0.12, C=0.8}
\end{figure}

Upon simulation of this model, we do not find evidence of superluminality. Commensurate with the MS case \cite{MS},  we observe in certain parameter choices the sonic horizon forming outside of the apparent horizon, as in the case of normal matter and  for $\mathcal{L}_{2}(X)$. We illustrate such results for a typical choice of parameters in figure \ref{lexpf}.    This offers an indication that the subluminality of this particular global-$t$ model is independent of the coordinate choice.  We also see the same behaviour as in the previous two cases with respect to the eventual merging of the horizons. For the parameter values chosen in figure \ref{lexpf}, for example, we see the two horizons as indistinguishable by P-G time t=0.449391824.    This is perhaps best illustrated in the map of null geodesics and horizons given in figure \ref{l3nullmap}.  We do not find Cauchy breakdown in this case, as expected by construction.  

For this model, we examined parameter pairings in the range of $A=0.02-0.13$ and $C=0.5-100.0$.  We note that, as discussed in \cite{MS}, the hyperbolicity requirement of equation (\ref{kefe}) prevents us from choosing $C<0.45$.  Although we encountered some numerical difficulties (explained below) that forced the program to halt before completion, these were unrelated to the physical or mathematical characteristics of the model, in contrast to the situation for $\mathcal{L}_{1}$ and $\mathcal{L}_{2}$.  With this in mind, we give the following overview of different types of behaviour found for this case.

\begin{enumerate}
\item{Subcritical collapse was observed for small $A$ for all choices of $C$.}
\item{For mid-range values of $A$ and small values of $C$, we observed a sonic horizon forming prior to the code halting, but not a luminal horizon. This is as expected from the construction of the model.}
\item{For mid-range values of $A$ and larger values of $C$, we observed neither sonic nor luminal horizons forming before the code halted.  This is due not to a different run time before halting (which was comparable at different values of $C$) but rather due to the fact that at higher $C$ (nearer the normal matter limit) the two horizon formations occur more closely together in time, so neither form before the code halts.}
\item{For large values of $A$ and small values of $C$, we observed subluminal behaviour such as that shown in figure \ref{lexpf}, with the sonic horizon forming outside the luminal horizon.  This is a natural extension of the behaviour at midsize $A$ and small $C$.  Here, since $A$ is larger, the horizons form earlier, and so both have time to form before the code halts.}
\item{For large values of $A$ and large values of $C$, we see both horizons forming before the code halts, but they are indistinguishable within our degree of precision.}
\end{enumerate}
  
We pause to comment on the numerical difficulties we encountered for this case. Because of the recursive nature of the definition of $X$ in P-G coordinates (see appendix A), we were forced to use a numerical solver to ascertain the value $X$ at a given time-step.  We chose a Newton-Raphson solution method, which had a very good rate of convergence. However  the tendency of $X$ to become unpredictably large near the origin after horizon formation  made it extremely difficult to find an initial guess  that would yield convergence for the entire duration and spatial extent of the simulation.  Hence we were forced to halt the code on those occasions that the Newton-Raphson method failed to converge. 
  
\section{Conclusions}

Through numerical simulations of the gravitational collapse of K-essence scalar fields in Painlev\'{e}-Gullstrand coordinates, we have been able to observe a number of interesting   phenomena.  For the  Lagrangian $\mathcal{L}_{1}$ we have numerically verified that sonic horizons form inside the luminal apparent horizon,  allowing superluminal perturbations to escape the black hole.  
For other choices of Lagrangian ($\mathcal{L}_{2}$ and $\mathcal{L}_{3}$), we see that collapse proceeds with the sonic horizon outside of the luminal horizon, indicating no possibility of superluminal escape. For all three Lagrangians we studied the two different horizons converge  relatively quickly, as the K-essence scalar field vacates the region of the horizons.  These results are fully commensurate with the MS case \cite{MS}, suggesting that they are quite general and coordinate independent.  

We can understand these features by examining the general form of the  sonic  horizon condition (\ref{shcond2}). Superluminal behaviour generically happens when the luminal horizon forms before the sonic horizon, yielding a region  in which (at least initially) light is trapped but not sound. As explained above, when ${\cal L}_{X}$ is positive this is definitely possible only when   ${\cal L}_{XX}$ is negative in the trapping region. For ${\cal L}_1$, we have:
\be 
{\cal L}_{1,X} = \frac{1}{\sqrt{1+2X/\alpha^2}} \qquad 
{\cal L}_{1,XX} = -\frac{1}{\alpha^2(1+2X/\alpha^2)^{3/2}}  
\label{F1}
\ee
whereas for ${\cal L}_2$:
\be
{\cal L}_{2,X} = \frac{1}{\sqrt{1-2X/\alpha^2}} \quad 
{\cal L}_{2,XX} = \frac{1}{\alpha^2(1-2X/\alpha^2)^{3/2}}  
\label{F2}
\ee
Moreover, for ${\cal L}_3$ all three quantities are manifestly positive. Thus, superluminal escape is in principle only possible for ${\cal L}_1$, and this is born out by our numerical calculations.

It is important to note that the above considerations are slicing independent. The presence, or not, of superluminal propagation inside a trapped surface of the luminal metric is coordinate invariant, although the precise location of the trapping horizons may depend on the slicing. 

The possibility of superluminal escape from a black hole can be considered in terms of the sonic and luminal event horizons of the black hole. These are defined as the last sound and light rays, respectively, to just barely escape from the black hole.  The sonic and luminal event horizons coincide along a line of constant $r$, say $r_h$, in the future after all the matter has fallen through. For $\mathcal{L}_1$ the sonic trapping horizon is to the future of the luminal trapping horizon, as  shown in Figure \ref{l1nullmap}. In this case, as one moves backward in time the sonic event horizon (red dashed line) ``peels off" towards small $r$ slightly faster than the luminal event horizon. Thus, there will be some K-essence perturbations just outside the sonic event horizon that will be able to escape to infinity from within the luminal event horizon.  A related question is whether or not sonic perturbations are able to escape from inside the luminal trapping horizon (light blue solid line). Although it is impossible to distinguish on the scale of Figure \ref{l1nullmap}, we have verified numerically that as the sonic event horizon separates from the trapping horizon, it does enter the luminal trapping horizon. For the parameters that we have considered the corresponding region of superluminal escape is very small. It is not clear whether or not there is a range of initial data that allows a region of superluminal escape that is large enough to affect the information loss problem. This is a question that is worthy of further investigation.

Note   that none of these three choices of Lagrangian are suitable for addressing the cosmic coincidence problem \cite{dynamic}.   For $\mathcal{L}_{1}$ we find that $X$ becomes negative in certain regions of space,  in contradiction with the assumed homogeneity of cosmology  \cite{MS}.
In the cases of $\mathcal{L}_{2}$ and $\mathcal{L}_{3}$, superluminality is not observed in either the MS or P-G simulations and so these models cannot be used to create the dynamical behaviour used to solve the problem of cosmic coincidence \cite{nogo}.  Any choice of $\mathcal{L}$ that solves this problem must incorporate superluminality.    These models are `K-essence' only in the sense of having non-linear kinetic energy.
 
As long as ${\cal L}_X$ and ${\cal L}_{XX}$ are regular, only ${\cal L}_1$ permits a breakdown of the Cauchy problem and this too was verified by the numerical calculations. However, in the case of ${\cal L}_2$, when $X$ is too large,  ${\cal L}_X$ and ${\cal L}_{XX}$ blow up. Our simulations (and those of \cite{MS}) verified that this behaviour does indeed occur in spherically symmetric collapse with ${\cal L}_2$.

For $\mathcal{L}_{1}$, we have observed  breakdown of the Cauchy problem.  The smaller the value of $\alpha$, the earlier this breakdown occurs during the simulation.  This indicates that the constant time surfaces in P-G coordinates fail to continue to be Cauchy surfaces for either the background metric or the effective metric of K-essence, a phenomenon also  observed for the maximal slicing time coordinate \cite{MS}.  This provides a compelling (albeit inconclusive) indication of the coordinate independence of this generic breakdown of the Cauchy problem for these choices of $\mathcal{L}$.  Whether or not avoidance of Cauchy breakdown is necessarily connected to subluminality remains an open question; it remains to be seen whether another time coordinate may offer a well-posed Cauchy problem throughout for such choices of $\mathcal{L}$.    If no time coordinate exists that ensures  the Cauchy problem is well-posed throughout collapse for both the background and effective metric, this would invalidate the predictive power of such models \cite{escape}, and eliminate the possibility of  the physically sensible escape of  information from a black hole.

\section*{Acknowledgements}
 We thank David Garfinkle and Ryo Saotome for helpful discussions and correspondence. We are grateful to SHARCNET for providing computer resources. This work was supported by the Natural Sciences and Engineering Research Council of Canada.

\section*{Appendix A}

In the following, where there is potential for confusion we denote 2 dimensional quantities with a hat and leave 4 dimensional quantities as they are.   Since we are considering spherically symmetric collapse, we dimensionally reduce the metric from 4 dimensions to 2 via
\be\label{met2d}
d{s}^2 = \frac{1}{j(\phi)}\hat{g}_{\mu\nu}dx^\mu dx^\nu + r^2\left(d\theta^2 + \sin^2(\theta)d\phi^2\right)
\ee
where $\phi = r^2/(4l^2)$ and $j(\phi) = \sqrt{\phi} = r/(2l)$, with $l$ an arbitrary length scale.  
Note that in the following $\hat{g}_{\mu\nu}$ and associated quantities refer to a 2-dimensional metric. The action (\ref{action}) becomes  
\be
\frac{1}{2\hat{G}}\int d^2x \sqrt{-\hat{g}}\left(\phi \hat{R} +\frac{V(\phi)}{l^2}\right)+ \int d^2x\sqrt{-\hat{g}} \frac{4\pi r^2}{j(\phi)} {\cal L}(j{\hat{X}})
\label{eq:dilaton action}
\ee
where $2\hat{G}= {G}/l^2$, $V(\phi) = 1/(2\sqrt{\phi})$, and
\be
\hat{X}= \frac{{X}}{j}= - \frac{1}{2}\hat{g}^{\mu\nu}\partial_\mu\psi\partial_\nu\psi
\ee

Employing an ADM decomposition of the 2-dimensional metric in (\ref{met2d}) 
\be
d\hat{s}^2 = e^{2\rho}\left(-\sigma^2 dt^2 + (dx+Ndt)^2\right)
\label{adm}
\ee
yields after some algebraic manipulation the Hamiltonian 
\be
H=\int dx \left( \frac{\sigma e^{2\rho}}{l\phi'}\left(-{\cal M}' + 4\pi r^2 \rho_M\right)
+\left(N-\frac{\sigma \hat{G} \Pi_\rho}{\phi'}\right){\cal F}\right)
\ee
where ${\cal M}$ is the Hamiltonian version of the Misner-Sharpe mass function 
\be
\frac{2\hat{G} {\cal M}}{l} \equiv e^{-2\rho}\left((\hat{G}\Pi_\rho)^2-(\phi')^2\right)+ \frac{j(\phi)}{l^2}
\label{eq:misner sharpe}  
\ee
and we define the mass energy density $\rho_M$ via
\be
4\pi r^2 \rho_M = \frac{l\phi'}{e^{2\rho}}\left(\frac{\Pi_\psi^2}{4\pi r^2 {\cal L}_{{X}}}- \frac{4\pi r^2 e^{2\rho}}{j(\phi)}{\cal L}({X})\right) + \frac{l\hat{G}\Pi_\rho}{e^{2\rho}} \Pi_\psi \psi' \, .
\label{eq:energy density}
\ee
The canonical momenta are given by
\bea
\Pi_\phi &=& \frac{1}{\sigma \hat{G}}(\rho'N+N'-\dot{\rho}) \\
\Pi_\rho &=& \frac{1}{\sigma \hat{G}}(\phi'N-\dot{\phi}) \\
\Pi_\psi &=& \frac{4\pi r^2}{\sigma}{\cal L}_{{X}}(\dot{\psi}-N\psi')
\eea
and we have defined
\be
{\cal F} =\rho'\Pi_\rho-\Pi_\rho'+\phi'\Pi_\phi + \Pi_\psi \psi'
\ee
It is important to note that the lapse, $\sigma$, and shift, $N$, in the above parametrization are also the lapse and shift functions of the physical 4-metric as defined in (\ref{met2d}).

Choosing the $r$  as the spatial coordinate implies both the consistency condition $\dot{\phi}=0$ and  $l\phi'=j(\phi)$.
Using this we obtain the shift in terms of the lapse:
\be
N=\frac{\sigma\hat{G} \Pi_\rho}{\phi'}
\label{ngen}
\ee
Setting the diffeomorphism constraints strongly to zero, we eliminate $\Pi_\phi$, yielding 
\be
H=\int dx \frac{\sigma e^{2\rho}}{j(\phi)} \left(-{\cal M}'+4\pi r^2 \rho_M\right)
  +\int dx (\sigma \frac{e^{2\rho}}{j} {\cal M})'
\label{eq:final hamiltonian}
\ee
for the partially reduced Hamiltonian upon including an appropriate boundary term.  This partial coordinate choice allows many different fully gauge fixed theories, including Schwarzschild coordinates, Painlev\'e-Gullstrand (P-G) coordinates and maximal slicings, among others.

Specializing to  P-G coordinates entails fixing  ${g}_{rr}=1$, or $e^{2\rho}=j(\phi)$. The consistency condition $\dot{\rho}=0$ fixes
the lapse via 
\be
\frac{\sigma'}{\sigma}=-\frac{l\hat{G}\Pi_\psi\psi'}{\sqrt{2l\hat{G}{\cal M}j(\phi)}}
\label{eq:sigma}
\ee
determining $\rho$ and $\sigma$. Making use of the above conditions with eq (\ref{eq:misner sharpe}) yields
\be
\frac{2\hat{G}{\cal M}}{l}=\frac{(\hat{G}\Pi_\rho)^2}{j(\phi)}
\label{mass1}
\ee
with $\Pi_\rho$  determined by the remaining Hamiltonian constraint.

Finally,  the field equations for the matter field independent variables $\psi$ and $\Pi_\psi$ from the Hamiltonian (\ref{eq:final hamiltonian}) 
yields after some manipulation
\bea\label{psidot}
\dot{\psi}&=&\sigma\left[\frac{\Pi_\psi}{4\pi r^2 {\cal L}_{{X}}}+\sqrt{\frac{2\hat{G}l{\cal M}}{j(\phi)}}\psi'\right] \\
\dot{\Pi}_\psi&=&\left[\sigma\left(4\pi r^2 {\cal L}_{{X}}\psi'+\sqrt{\frac{2\hat{G}l{\cal M}}{j(\phi)}}\Pi_\psi\right)\right]'
\label{Pipsidot}
\eea
The quantities 
${\cal M}$ and $\sigma$ as determined by the Hamiltonian constraint (\ref{mass1}) 
\be
-{\cal M}' + 4\pi r^2 \rho_M \sim 0
\ee
and the consistency condition (\ref{eq:sigma}), giving
\be
\frac{\sigma'}{\sigma}=-\frac{l\hat{G}\Pi_\psi\psi'}{\sqrt{2l\hat{G}{\cal M}j(\phi)}}
\ee
with $\rho_M$ given by 
\be
4 \pi r^2 \rho_M=l\left(\frac{\Pi_\psi^2}{4\pi r^2 {\cal L}_{\overline{X}}}- 4\pi r^2 {\cal L}(\overline{X})\right)+ \sqrt{\frac{2 \hat{G} {\cal M}l}{j(\phi)}} \Pi_\psi \psi'
\label{eq:PG energy density}
\ee
Recalling that $2\hat{G}= {G}/l^2$, we note that the dynamical equations for the K-essence field and its conjugate momentum may be expressed in terms of four dimensional quantities only. 

As an example, for the choice (\ref{sh2}) we have
\be
{\cal L}_{{X}} = \frac{1}{\sqrt{1-\frac{2{X}}{\alpha^2}}}\qquad 
{\cal L}_{{X}{X}}=\frac{1}{\alpha^2}\left(1-\frac{2{X}}{\alpha^2}\right)^{-3/2}
\ee
In this case we can solve explicitly for ${X}$ in terms of the phase space variables 
\be 
{X} = \frac{1}{2}\left(\frac{\Pi_\psi^2}{(4\pi r)^2 {\cal L}^2_{{X}}}-(\psi')^2\right) 
 = -\frac{\alpha^2}{2}\left[\frac{(4 \pi r^2 \psi^\prime)^2 - \Pi_\psi^2}{(4 \pi r^2 \alpha)^2 + \Pi_\psi^2}\right]
\ee 
in P-G coordinates.

For the purposes of our numerical simulations, we set the arbitrary length scale $l$ to $l=1$.  Choosing the initial configuration  for  $\psi$ to be that given in equation (\ref{initpsi}), we first construct a fixed, spherically symmetric spatial grid in which to evolve the system, by specifying the number of spatial steps (we use 600), the minimum space between steps, and the maximum space between steps.  Near the origin we require a finer grid. We therefore initialize the spatial grid with the minimum space between steps for a specified number of first steps nearest to the origin (we use 100 steps).  Once we are 100 steps from the origin, the spatial grid is gradually expanded so that the space between the grid points is at a maximum once we reach grid point 600.  We select a minimum spacing of $1.0 \times 10^{-3}$ and a maximum spacing of $1.0 \times 10^{-2}$.  After specifiying the initial configuration of the spatial grid, we then initialize the configuration of $\psi$ as specified above and we initialize the conjugate momentum $\Pi_{\psi}$ to be identically 0.  

After these initial conditions have been specified, we enter a loop, which proceeds at each timestep until a predetermined and specfied maximal time is reached after which we anticipate no interesting evolution of the system.  Note that while the spatial grid is constant at each timestep and only specified once, the timestep is not constant.  It is determined during each execution of the loop.

Inside the loop, we first calculate the necessary derivatives of $\psi$ and $\Pi_{\psi}$ using finite difference methods.  We then calculate $\mathcal{L}$ and its derivatives, as well as $X$ and its first dervative in r.  Next, we determine $\mathcal{M}$ and $\sigma$ on the spatial grid at the current timestep using a fourth order Runge-Kutta method.  We then employ the condition that determines the location of the  luminal apparent horizon and the sonic apparent horizon, as well as checking for Cauchy breakdown.

At this point in the loop, data is written to file on selected executions of the loop.  We then calculate the next timestep, according to the speed of a local ingoing geodesic.  Finally, we evolve $\psi$ and $\Pi_{\psi}$ in time using another fourth order Runge-Kutta method, and increase our total time by the calculated timestep.  The loop then restarts.

The loop continues to run until the total time has reached some predetermined and declared value as mentioned above.   Alternatively, the program may cease to run due to a check performed on the values of $\mathcal{L}$ and its derivatives each time they are computed.  If these values become divergent or imaginary, the  program  halts.  This corresponds in the case of $\mathcal{L}_{1}$ to Cauchy breakdown, and in the case of $\mathcal{L}_{2}$ to the divergence of $\mathcal{L}_{2 X}$.  The  program may also halt for $\mathcal{L}_{3}$ due to a check on the convergence of the Newton-Raphson solver.

\end{document}